\documentclass{article}

\usepackage{emulateapj}
\usepackage{apjfonts}
\usepackage{psfig}
\usepackage[american]{babel}

\newcommand{\Msun}{\ensuremath{M_{\odot}}}
\newcommand{\Zsun}{\ensuremath{Z_{\odot}}}

\newcommand{\Ae}{$\alpha$-enhancement}
\newcommand{\Hb}{\ensuremath{{\rm H}\beta}}
\newcommand{\Zmaj}{\ensuremath{Z_{\rm maj}}}
\newcommand{\Zsub}{\ensuremath{Z_{\rm sub}}}
\newcommand{\Zssp}{\ensuremath{Z_{\rm SSP}}}
\newcommand{\tmaj}{\ensuremath{t_{\rm maj}}}

\newcommand{\BK}{\ensuremath{B\!-\!K}}
\newcommand{\UV}{\ensuremath{U\!-\!V}}
\newcommand{\UVV}{\ensuremath{1500\!-\!V}}

\submitted{Accepted for publication in the Astrophysical Journal}

\lefthead{CLAUDIA MARASTON AND DANIEL THOMAS}
\righthead{STRONG BALMER LINES IN OLD STELLAR POPULATIONS}

\begin{document}

\title{Strong Balmer lines in old stellar populations:\\ 
No need for young ages in ellipticals?}

\author{Claudia Maraston and Daniel Thomas}
\affil{Universit\"ats-Sternwarte M\"unchen, Scheinerstr.\ 1,
D-81679 M\"unchen, Germany}
\authoraddr{Scheinerstr.~1, D-81679 M{\"u}nchen, Germany; maraston{\@@}usm.uni-muenchen.de}

\begin{abstract}
Comparing models of Simple Stellar Populations (SSP) with observed line
strengths generally provides a tool to break the age-metallicity degeneracy
in elliptical galaxies. Due to the wide range of Balmer line strengths
observed, ellipticals have been interpreted to exhibit an appreciable
scatter in age. In this paper, we analyze Composite Stellar Population
models with a simple mix of an old metal-rich and an old metal-poor
component. We show that these models simultaneously produce strong Balmer
lines and strong metallic lines without invoking a young population. The key
to this result is that our models are based on SSPs that better match the
steep increase of \Hb\ in metal-poor globular clusters than models in the
literature. Hence, the scatter of \Hb\ observed in cluster and luminous
field elliptical galaxies can be explained by a spread in the metallicity of
{\em old} stellar populations. We check our model with respect to the
so-called G-dwarf problem in ellipticals. For a galaxy subsample covering a
large range in \UVV\ colors we demonstrate that the addition of an old
metal-poor subcomponent does not invalidate other observational constraints
like colors and the flux in the mid-UV.
\end{abstract}


\section{Introduction}
More than 20 years ago it has been recognized that the modeling of the
spectral energy distribution of ellipticals is affected by an ambiguity in
age and metallicity (Faber 1972; O'Connell 1976)\nocite{Fa72,Oc76}, which
has turned out to be a general complication in population synthesis (Renzini
1986)\nocite{Re86}. However, considering Simple Stellar Populations (SSP) in
the two-parameter space of Balmer and metallic lines, the age-metallicity
degeneracy can be broken (Gonz\'alez 1993; Worthey 1994)\nocite{G93,Wo94}.
The major reason for the success of this strategy is that the Balmer line
strengths of SSPs are predominantly age sensitive at metallicities above
$\sim 1/3~\Zsun$ that are supposed to be the only relevant for elliptical
galaxies. Strong \Hb\ lines are thus taken as an indication for young
(intermediate-age) populations, the observed scatter is interpreted as a
considerable spread in age (e.g., Faber et al.\ 1995)\nocite{Fetal95}.

An alternative path to obtain blue stars and hence strong Balmer lines is to
consider old {\em metal-poor} populations. In this paper we follow this
approach and compute composite stellar populations that contain an old
metal-poor subcomponent. The principal focus is to check if a combination of
only old populations can reproduce strong \Hb\ without invalidating further
constraints for ellipticals like metallic indices, colors and spectral energy
distributions.

The paper is organized as follows. In Section~\ref{calsec} we calibrate the
SSP model indices and spectral energy distributions on globular
cluster data. The composite models and their application to galaxy
data are presented in Section~\ref{resultssec}. In
Sections~\ref{discusssec} and~\ref{sumsec} we discuss and summarize
the results.

\section{Simple stellar population models}
\label{calsec}
We compute a new set of SSP models based on the population synthesis
presented in Maraston (1998)\nocite{Ma98}, in which the fuel consumption
theorem (Renzini \& Buzzoni 1986)\nocite{RB86} is adopted to evaluate the
energetics of the post main sequence phases. The new SSP models cover the
metallicities $-2.25<{\rm [Fe/H]}<0.5$ and ages $10^7-2\cdot 10^{10}$~yr.
The input stellar tracks are taken from Bono et al.\ (1997)\nocite{Betal97}
and S.~Cassisi (1999, private communication). These new models will be
discussed in detail in a future paper (C.\ Maraston, in preparation). The
calibration of the model colors on globular clusters is presented in
Maraston (1998, 2000)\nocite{Ma98,Ma00}. Here we show the calibration of
indices and spectral energy distributions of the SSP models that are
relevant for the present study.

\subsection{Spectral line indices}
\begin{figure*}[ht!]
\begin{center}
\begin{minipage}{0.5\textwidth}
\psfig{figure=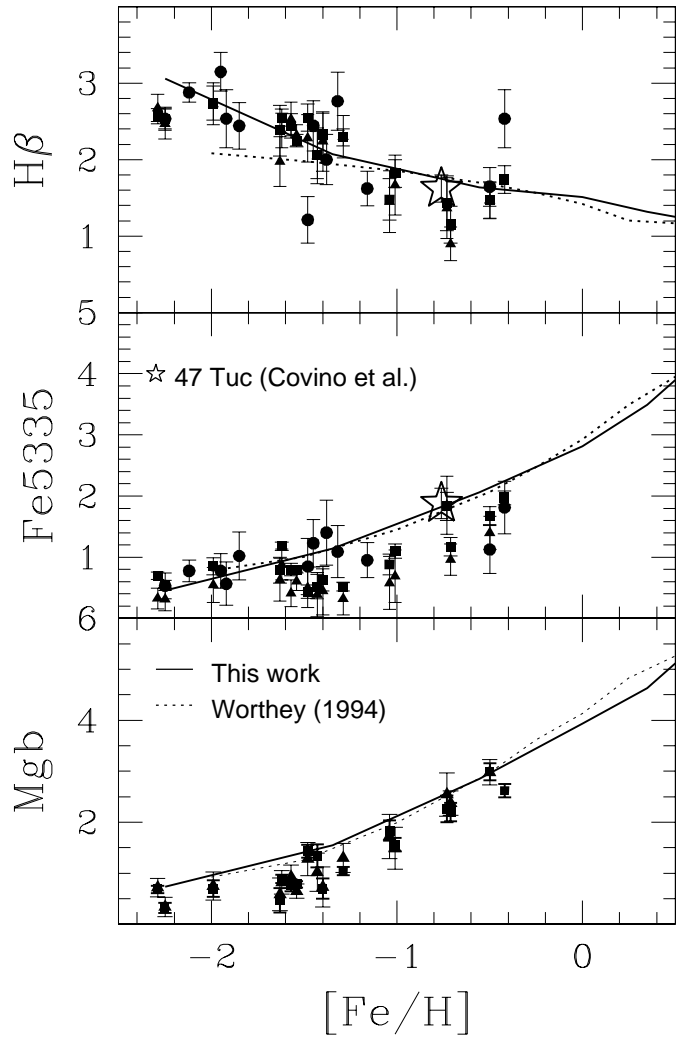,width=\textwidth}
\end{minipage}
\end{center}
\caption{Calibration of the SSP models on galactic globular clusters. 
Data are from Burstein et al.\ 1984 (squares), Covino, Galletti \&
Pasinetti 1995 (circles), and Trager 1998 (triangles). [Fe/H] 
(Zinn \& West (1984) scale) is taken from Harris (1996). Solid lines are our
SSPs for age $t=15$ Gyr. Worthey (1994) SSPs for $t=17$ Gyr are shown
as dotted lines.\label{fehbssp}}
\nocite{Wo94,Ha96}
\nocite{ZW84,Betal84,CGP95,Tra98}
\end{figure*}

Synthetic line indices for SSPs are computed using the fitting functions
from Worthey et al.\ (1994)\nocite{Wetal94}. Fig.~\ref{fehbssp} shows \Hb,
Fe5335, and Mgb as a function of [Fe/H] for galactic globular clusters from
various data sets (see the caption). Our SSP models of fixed age (15~Gyr)
and various metallicities are plotted as solid lines. Dotted lines are the
models from Worthey (1994)\nocite{Wo94} for $t=17$~Gyr.

As metal-poor stars (${\rm [Fe/H]}\la -1.3$) spend their horizontal branch
lifetimes at temperatures $T_{\rm eff}\ga 5500$~K, \Hb\ is steeply rising
with decreasing [Fe/H] (upper panel in Fig.~\ref{fehbssp}). This observed
feature is the key for the present analysis and is well reproduced by our
SSPs. The models from Worthey (1994)\nocite{Wo94}, instead, predict a
shallower trend, which implies that the very metal-poor globulars are
younger than the more metal-rich ones. The synthetic Fe5335 indices are in
good agreement with the Covino, Galletti
\& Pasinetti (1995)\nocite{CGP95} data, while for ${\rm [Fe/H]}\ga -1.5$
Burstein et al.\ (1984)\nocite{Betal84} and Trager (1998)\nocite{Tra98}
derive lower values. Mgb indices seem slightly overestimated by the models
at low metallicities.

Our synthetic metallic indices Fe5335, Mgb and the Balmer line \Hb\ for
${\rm [Fe/H]}\ga -1.5$ agree well with the models from Worthey (1994),
because both sets of models are based on the same fitting functions (Worthey
et al.\ 1994)\nocite{Wetal94}. The uncertainty introduced by the use of
different fitting functions is discussed in Maraston, Greggio \& Thomas
(1999) and L.\ Greggio \& C.\ Maraston (in preparation)\nocite{MGT99}. The
low \Hb\ values given by Worthey (1994)\nocite{Wo94} at very low
metallicities are likely due to the assumption of too red horizontal branch
morphologies in the models.

As discussed in Worthey et al.\ (1994)\nocite{Wetal94}, iron indices are
favorable as metallicity indicators, because the analysis with Mg-indices is
severely affected by \Ae\ (Tripicco \& Bell 1995)\nocite{TB95}: SSP models
based on solar abundance ratios predict -- for a given Fe index -- weaker Mg
indices than measured in elliptical galaxies (Worthey, Faber \& Gonz\'alez
1992; Greggio 1997)\nocite{WFG92,Gre97}. As at high metallicities the
fitting functions of Fe5270 are significantly more uncertain (Maraston et
al.\ 1999)\nocite{MGT99}, in the following analysis we focus on \Hb\ and
Fe5335. We additionally consider the color \BK\ as metallicity indicator
which turns out to be in good agreement with Fe5335, while Mgb
systematically implies higher metallicities (see Section~\ref{resultssec}).

Combining \Hb\ with Fe5335, for 47 Tuc we obtain a
`spectroscopic age' of 15~Gyr (Fig.~\ref{fehbssp}) in agreement with the age
derived from the color-magnitude diagram ($14\pm 1$~Gyr, Richer et al.\
1996)\nocite{Rietal96}. This example shows that the \Hb-Fe5335 plane is
successful in breaking the age-metallicity degeneracy, when applied to simple
systems like globular clusters. Note that the high resolution Balmer
index H$\gamma_{\rm HR}$, instead, leads to a `spectroscopic age' in
excess of $20$~Gyr (Gibson et al.\ 2000)\nocite{Getal99}.

\subsection{Spectral energy distributions}
\label{sedcalsec}
\begin{figure*}[ht!]
\psfig{figure=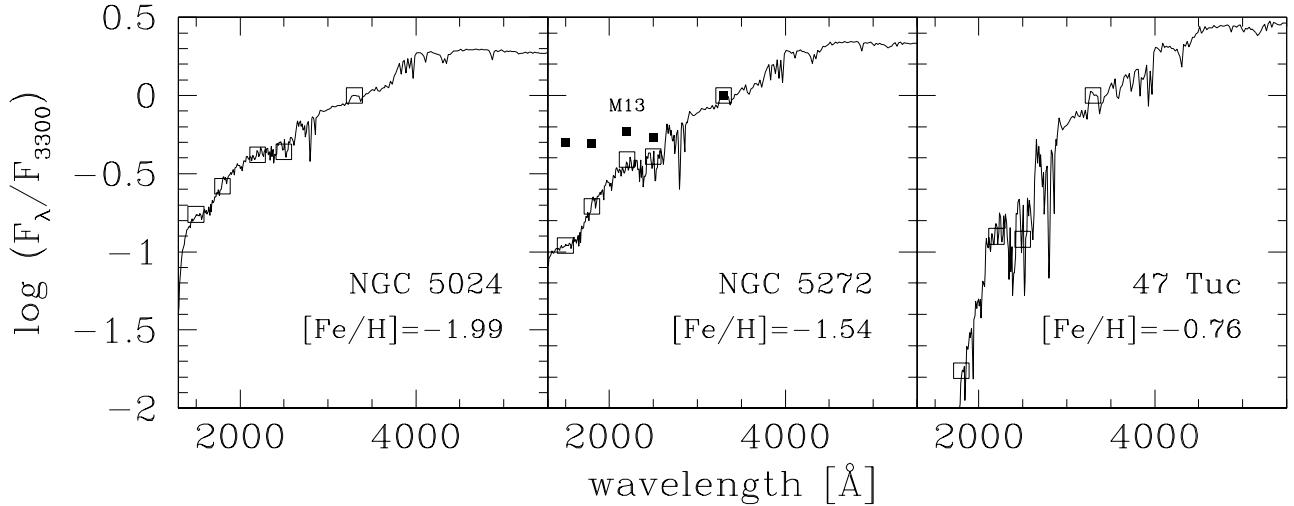,width=\linewidth}
\caption{Calibration of the synthetic SEDs in the UV on galactic
globular clusters. Data (squares) are taken from van Albada, de Boer \&
Dickens (1981), [Fe/H] (Zinn \& West (1984) scale) is adopted from Harris
(1996). The filled squares (central panel) show M13 that has an extreme blue
horizontal branch morphology. The solid lines are our model SEDs ($t=15$~Gyr)
with metallicities as indicated in the panels.
\label{sedcal}}
\nocite{vABD81,Ha96}
\end{figure*}
The evolutionary synthesis code of Maraston (1998)\nocite{Ma98} is updated
for the computation of the SSP spectral energy distributions (SEDs). The
spectral library of Lejeune, Cuisinier \& Buser (1998)\nocite{LCB98} is
adopted to describe the stellar spectra as functions of gravity, temperature
and metallicity.

The tightest constraint on the amount and the metallicity of metal-poor
stars in elliptical galaxies comes from the flux in the mid-UV
($2000-4000$~\AA). In Fig.~\ref{sedcal} we present the calibration of
synthetic metal-poor SEDs on galactic globular cluster data in the
wavelength range $\lambda=1500-3300$~\AA\ (van Albada, de Boer \& Dickens
1981)\nocite{vABD81}. The model spectra are in excellent agreement with the
data in the whole metallicity range. The increase of the flux shortward
3000~\AA\ with decreasing metallicity is due to a hotter main sequence
turn-off and a bluer horizontal branch. Note, however, that there are
globulars of intermediate metallicity ([Fe/H]$\sim-1.5$) that show
horizontal branches bluer than what it is expected with the canonical
mass-loss on the Red Giant Branch (van Albada et al.\ 1981)\nocite{vABD81}.
This case is shown in the central panel of Fig.~\ref{sedcal} for the
globulars M13 (NGC~6205), filled squares) and NGC~5272, which have the same
metallicity ${\rm [Fe/H]}\sim-1.54$ but different horizontal branch
morphologies. M13 contains extreme blue horizontal branch stars (`blue
tails'), and exhibits an excess of flux shortward $\sim 2000$~\AA\ (see
Section~\ref{m31sec}).

\begin{figure*}[ht!]
\psfig{figure=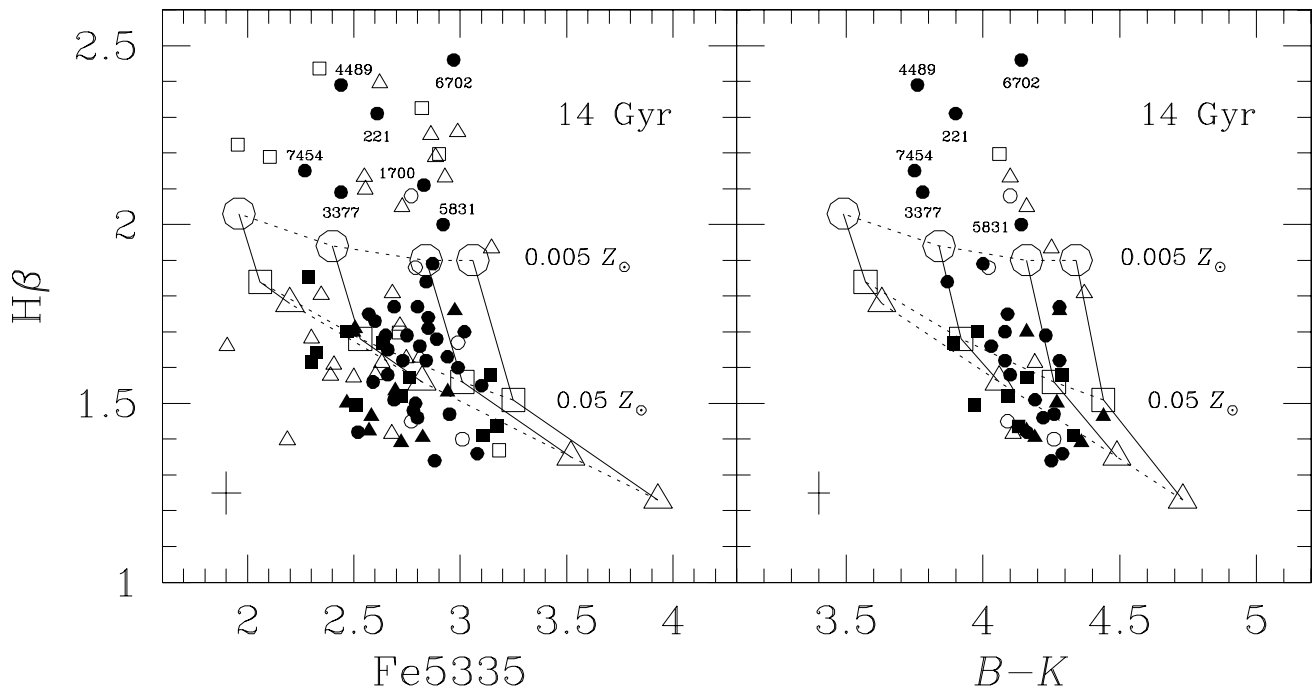,width=\linewidth}
\caption{\Hb\ vs.\ Fe5335 and \BK\ for elliptical (small filled symbols) and
lenticular (small open symbols) galaxies. Data are from Gonz\'alez 1993
(circles, Virgo and field, $R_{\rm e}/8$ aperture, labels indicate NGC
numbers), Kuntschner \& Davies 1998 (squares, Fornax), and Mehlert et al.\
2000 (triangles, Coma). \BK\ colors are taken from Pahre (1998). The error
bars denote average 1$\sigma$ errors. SSP models for $\Zssp=0.5,1,2,3~\Zsun$
are large open triangles, with \Zssp\ increasing from left to right. Composite
models for $\Zmaj=\Zssp$ and $\Zsub=0.05,0.005~\Zsun$ are indicated by large
open squares and large open circles. Solid and dotted lines denote fixed
\Zmaj\ and fixed \Zsub, respectively. The metal-poor subpopulation contributes
10 per cent by mass. The ages of the major component and the subcomponent are
14 Gyr and 15 Gyr, respectively.}
\label{fehb}
\nocite{G93,KD98,Pa98,Mehetal00}
\end{figure*}
As blue tails are preferentially found in denser and more concentrated
globulars, it is likely that the particular dynamical conditions in such
`second parameter' globular clusters lead to enhanced mass-loss and
extremely hot horizontal branch stars (Fusi Pecci et al.\
1993)\nocite{Fuetal93}, which may not apply to the average stellar field
population in galaxies. Since in early-type galaxies globular clusters
contribute less than 0.6 per cent to the total $V$-light (Ashman \& Zepf
1998)\nocite{AZ98}, we adopt the canonical mass loss $\eta=0.33$ for
[Fe/H]$\sim-1.5$ calibrated on `normal' globulars like NGC~5272 (Fusi Pecci
\& Renzini 1975)\nocite{FR75}.

\section{Results}
\label{resultssec}
In the following we compare old ($>12$~Gyr) composite stellar
populations with data of elliptical galaxies. The models consist of
two components: a major metal-rich population ($Z=\Zmaj$) and a metal-poor
subpopulation ($Z=\Zsub$). More complex models with a continuum of 
metallicities will be subject of a future paper.
\subsection{Indices and colors}
Fig.~\ref{fehb} displays \Hb\ vs.\ Fe5335 (left-hand panel) and \Hb\ vs.\ \BK\
(right-hand panel) for composite models in which the ages of the major 
population and of the metal-poor subcomponent are fixed to 14~Gyr and
15~Gyr, respectively. The contribution from the low-metallicity
population is 10 per cent by mass.

The grid in Fig.~\ref{fehb} shows models for various metallicities of the
two components. Solid lines connect models with constant metallicity of the
major component $\Zmaj=0.5,1,2,3~\Zsun$ (from left to right). The
metallicity of the subcomponent is $\Zsub=0.05$ (large open squares) and
$\Zsub=0.005~\Zsun$ (large open circles). For comparison, simple stellar
populations with metallicities $Z=\Zmaj$ and $t=\tmaj$ are shown as large
open triangles. The dotted (horizontal) lines mark models of constant \Zsub.

The perturbation of metal-rich simple stellar populations with metal-poor
stars leads to lower Fe5335 line strengths, bluer \BK\ colors, and stronger
Balmer lines. The models with $\Zsub=0.005~\Zsun$ reach \Hb\ $\sim 2$~\AA.
Observed line strengths (Gonz\'alez 1993; Kuntschner \& Davies 1998; Mehlert
et al.\ 2000)\nocite{G93,KD98,Mehetal00} and colors (Pahre 1998)\nocite{Pa98}
of elliptical galaxies and lenticular galaxies are shown as small filled and
small open symbols, respectively. The present set of composite models match
the area that is covered by the majority of the galaxy data. Note that the
objects exhibiting $\Hb>2$~\AA\ are either classified as lenticular galaxies
(small open symbols) or they are intermediate-mass {\em field} ellipticals
(NGC~1700, NGC~3377, NGC~5831, NGC~6702, NGC~7454) or dwarf ellipticals (M32
(NGC~221), NGC~4489) from the Gonz\'alez (1993)\nocite{G93} sample. All {\em
luminous field ellipticals} and all {\em cluster ellipticals} (Coma, Fornax,
and Virgo) in Fig.~\ref{fehb}, instead, can be modeled with the old
composite populations introduced above. It should be emphasized that the
bulk of ellipticals scatter around $\Hb\sim 1.6$~\AA, so that less than 10
per cent contribution from metal-poor stars is required in most cases. The
observed range $\Hb\sim 1.3-2.0$\ is then due to a spread in the metallicity
and/or the weight of the metal-poor subcomponent. The analysis with simple
stellar population models, instead, yields an {\em age}-spread of $\sim
7-15$~Gyr.

\begin{figure*}[ht!]
\begin{center}
\begin{minipage}{0.5\textwidth}
\psfig{figure=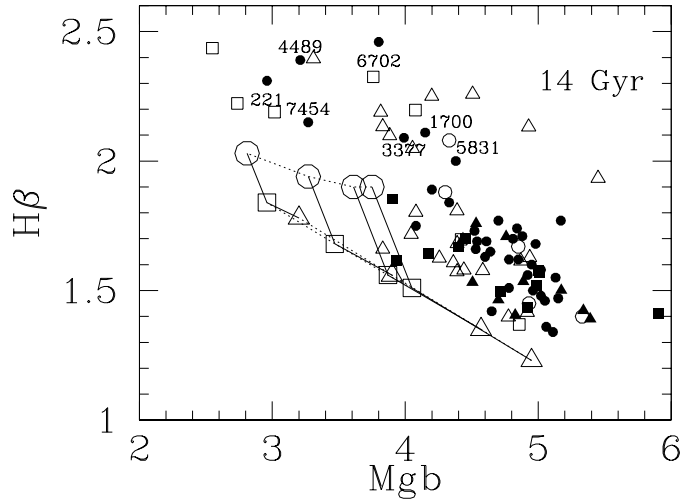,width=\textwidth}
\caption{\Hb\ vs.\ Mgb. Symbols and linestyles like in Fig.~\ref{fehb}.}
\label{mghb}
\end{minipage}
\end{center}
\end{figure*}

Another striking feature of Fig.~\ref{fehb} is that the positions of the
galaxy data relative to each other are the same in the Fe5335 and the \BK\
diagrams. Moreover, the locations of the model grids relative to the data
points in the two panels are in good agreement, so that both the metallic
index Fe5335 and the color \BK\ consistently constrain the metallicity range
for the major component to $1-3~\Zsun$. This strongly reinforces the use of
Fe5335 as metallicity indicator (see Section~\ref{calsec}). In
Fig.~\ref{mghb} we show the same data and the same model grid in the \Hb-Mgb
plane. As discussed in Section 2.1, Mg lines of ellipticals are generally
stronger than any SSP model. This feature is attributed to an enhancement of
Mg in luminous ellipticals. As a consequence the metallicities derived from
the Mg-indices are substantially higher and provoke severe inconsistencies
with observed colors (see also Saglia et al.\ 2000)\nocite{Saetal00}. As the
latter are much less affected by abundance ratio anomalies, the $\Hb$-\BK\
plane represents the more reliable constraint on stellar population
models.

Note that the \Hb\ line index additionally suffers from uncertainties in the
correction for contamination by emission (Gonz\'alez 1993)\nocite{G93},
which can lead to an underestimation of the \Hb\ line strength. This effect
may explain the low values $\Hb\la 1.5$~\AA\ that fall below the model grid
in Fig.~\ref{fehb}.

\subsection{Constraints from the mid-UV}
\label{uvsec}

The fluxes radiated by metal-poor stellar populations are high in the mid-UV
($2000-4000$~\AA), and decrease shortward 2000~\AA\ (Fig.~\ref{sedcal}). The
so-called UV-upturn below 2000~\AA\ observed in elliptical galaxies
(Burstein et al.\ 1988)\nocite{Betal88}, instead, is most likely due to late
evolutionary phases of old metal-rich populations (Greggio \& Renzini 1990;
Dorman, O'Connell \& Rood 1995; Yi, Demarque \& Oemler 1998; Greggio \&
Renzini 1999)\nocite{DOR95,YDO98,GR90,GR99}. The amount of metal-poor stars
in elliptical galaxies is therefore tightly constrained by the flux in the
mid-UV around 2500~\AA.

A model based on the closed-box metallicity distribution is not compatible
with the mid-UV flux of the elliptical galaxies NGC~4649 and
NGC~1404 (Bressan, Chiosi \& Fagotto 1994) and of the
M31 bulge (Worthey, Dorman \& Jones 1996)\nocite{BCF94,WDJ96},
which is referred to as the G-dwarf problem for ellipticals. On the other
hand, Lotz, Ferguson \& Bohlin (2000)\nocite{LFB00} find that the spectra of
ellipticals in the mid-UV even require the addition of a metal-poor component.
In the following we investigate if the proportions of metal-poor stars
required to explain the \Hb\ line strengths are consistent with the fluxes
observed in the mid-UV.

The objects with the strongest \Hb\ ($\sim 1.65$) that we obtain from cross
correlating the samples of Gonz\'alez (1993)\nocite{G93}, Kuntschner \&
Davies (1998)\nocite{KD98}, and Mehlert et al.\ (2000)\nocite{Mehetal00}
with the Burstein et al.\ (1988)\nocite{Betal88} sample (`quiescent
objects') are M31 (NGC~224), NGC~4472, and NGC~3379. We additionally
consider NGC~4649, which exhibits a low $\Hb=1.40$. These objects cover a
large range in \UVV\ colors, NGC~4649 having the strongest UV-upturn. In the
following we check if the composite models for these specific galaxies are
in agreement with their SEDs. In order to constrain the flux from the
metal-poor population in the mid-UV, it is necessary to model the UV-upturn.
Following the description by Greggio \& Renzini (1990)\nocite{GR90}, we
include hot ($T\sim 25,000$~K) late evolutionary phases in the metal-rich
components, such that the observed UV rising branches are reproduced.

The main model parameters for each galaxy are listed in
Table~\ref{modeltab}. Col.~(3) specifies the mass fractions (per cent) of
the two components. Note that less than 10 per cent is required in all
cases. The metallicities, ages, indices (\Hb, Fe5335, Mgb), and colors (\BK,
\UV) of the two populations are given in Cols.~$(4)-(10)$. The fuel F$_{\rm
UV}$ (in \Msun) in late evolutionary phases required to reproduce the
observed UV-upturn is given in Col.~(11).

\begin{deluxetable}{llccccccccc}
\tablecaption{Model parameters}
\tablewidth{0pt}
\tablehead{\colhead{galaxy} & \colhead{comp.} & \colhead{mass (\%)} & \colhead{$Z$ (\Zsun)} & \colhead{$t$ (Gyr)} & 
\colhead{\Hb} & \colhead{Fe5335} & \colhead{Mgb} & \colhead{\BK} & \colhead{\UV} & \colhead{F$_{\rm UV}$ (\Msun)} }
\tablecolumns{11}
\startdata
M31     & major & 97 & 1.750  &  12  & 1.51 & 3.20 & 4.38 & 4.34 & 1.73 & 0.003   \\
        & sub   &  3 & 0.005  &  15  & 3.09 & 0.53 & 0.78 & 2.64 & 0.53 & \nodata \\
N4472   & major & 94 & 2.000  &  12  & 1.48 & 3.32 & 4.54 & 4.42 & 1.81 & 0.003   \\
        & sub   &  6 & 0.030  &  15  & 2.33 & 1.02 & 1.38 & 2.91 & 0.72 & \nodata \\
N3379   & major & 99 & 1.000  &  14  & 1.56 & 2.82 & 3.90 & 4.06 & 1.59 & 0.002   \\
        & sub   &  1 & 0.005  &  15  & 3.09 & 0.53 & 0.78 & 2.64 & 0.53 & \nodata \\
N4649   & major & 90 & 1.550  &  15  & 1.38 & 3.21 & 4.41 & 4.36 & 1.73 & 0.020   \\
        & sub   & 10 & 0.300  &  15  & 1.69 & 2.08 & 2.87 & 3.53 & 1.15 & \nodata \\
\enddata
\label{modeltab}
\end{deluxetable}
\begin{deluxetable}{lccccc}
\tablecaption{Flux contributions}
\tablewidth{0pt}
\tablehead{\colhead{galaxy} & \colhead{\Hb} & \colhead{5550\AA} & \colhead{3600\AA} & \colhead{2500\AA} & 
\colhead{1500\AA}}
\tablecolumns{6}
\startdata
M31    &    9     &    6     &   17     &  61     &  28       \\  
N4472  &   16     &   12     &   30     &  70     &  16       \\  
N3379  &    4     &    2     &    5     &  33     &  14       \\
N4649  &   19     &   17     &   27     &  16     &   0       \\
\enddata
\tablecomments{The Table gives the relative contributions from the
subcomponent in per cent to \Hb\ (Col.~(2)) and to the flux at the wavelengths
specified in Cols.~$(3)-(6)$.}
\label{conttab}
\end{deluxetable}
\begin{deluxetable}{lcccccccc}
\tablecaption{Results}
\tablewidth{0pt}
\tablehead{ & \multicolumn{2}{c}{\Hb} & \multicolumn{2}{c}{Fe5335} & \multicolumn{2}{c}{\BK} & \multicolumn{2}{c}{\UV}\\
\colhead{galaxy} & \colhead{model} & \colhead{observed} & \colhead{model} & \colhead{observedd} & \colhead{model} & \colhead{observed} & \colhead{model} & \colhead{observed}}
\tablecolumns{9}
\startdata
M31    &  1.65    & $1.67\pm 0.07$ & 3.04  & $2.99\pm 0.06$       &   4.26   & \nodata &  1.61 &  \nodata   \\
N4472  &  1.62    & $1.62\pm 0.06$ & 3.04  & $2.84\pm 0.08$       &   4.28   & $4.28$  &  1.59 &  $1.57$    \\
N3379  &  1.62    & $1.62\pm 0.05$ & 2.77  & $2.73\pm 0.04$       &   4.05   & $4.08$  &  1.55 &  $1.52$    \\
N4649  &  1.44    & $1.40\pm 0.05$ & 3.01  & $3.01\pm 0.05$       &   4.24   & $4.26$  &  1.61 &  $1.61$    \\
\enddata
\tablecomments{Indices and colors of the models given in
Table~\ref{modeltab}. Index data ($1\sigma$ errors) are from Gonz\'alez
(1993), \BK\ from Pahre (1998), and \UV\ from Trager (1998).}
\label{resulttab}
\end{deluxetable}

In Table~\ref{conttab} we present the relative contributions to \Hb\ and to
the total fluxes at the wavelengths 5550 ($V$-band), 3600 ($U$-band), 2500,
and 1500~\AA. The weight of the metal-poor component increases with
decreasing $\lambda$, with a maximum contribution in the mid-UV ($\sim
2500$~\AA) as the rising branch shortward 2000~\AA\ is produced by the
metal-rich component. It is important to emphasize that in the model
proposed here, the Balmer line strengths and the UV upturn do not have a
common origin: we obtain strong \Hb\ with {\em metal-poor} (hot) horizontal
branch stars, while blue \UVV\ colors are due to extremely hot old {\em
metal-rich} stars.  The trend of decreasing \Hb\ with increasing Mg$_2$ and
decreasing \UVV\ (Burstein et al.\ 1988)\nocite{Betal88} is therefore not
affected.

In Table~\ref{resulttab} we compare the resulting indices and colors of the
composed models with the observed values. The indices \Hb, Fe5335 and the
colors \BK, \UV\ are in good agreement with observations.

\begin{figure*}[ht]
\begin{center}
\begin{minipage}{0.75\linewidth}
\psfig{figure=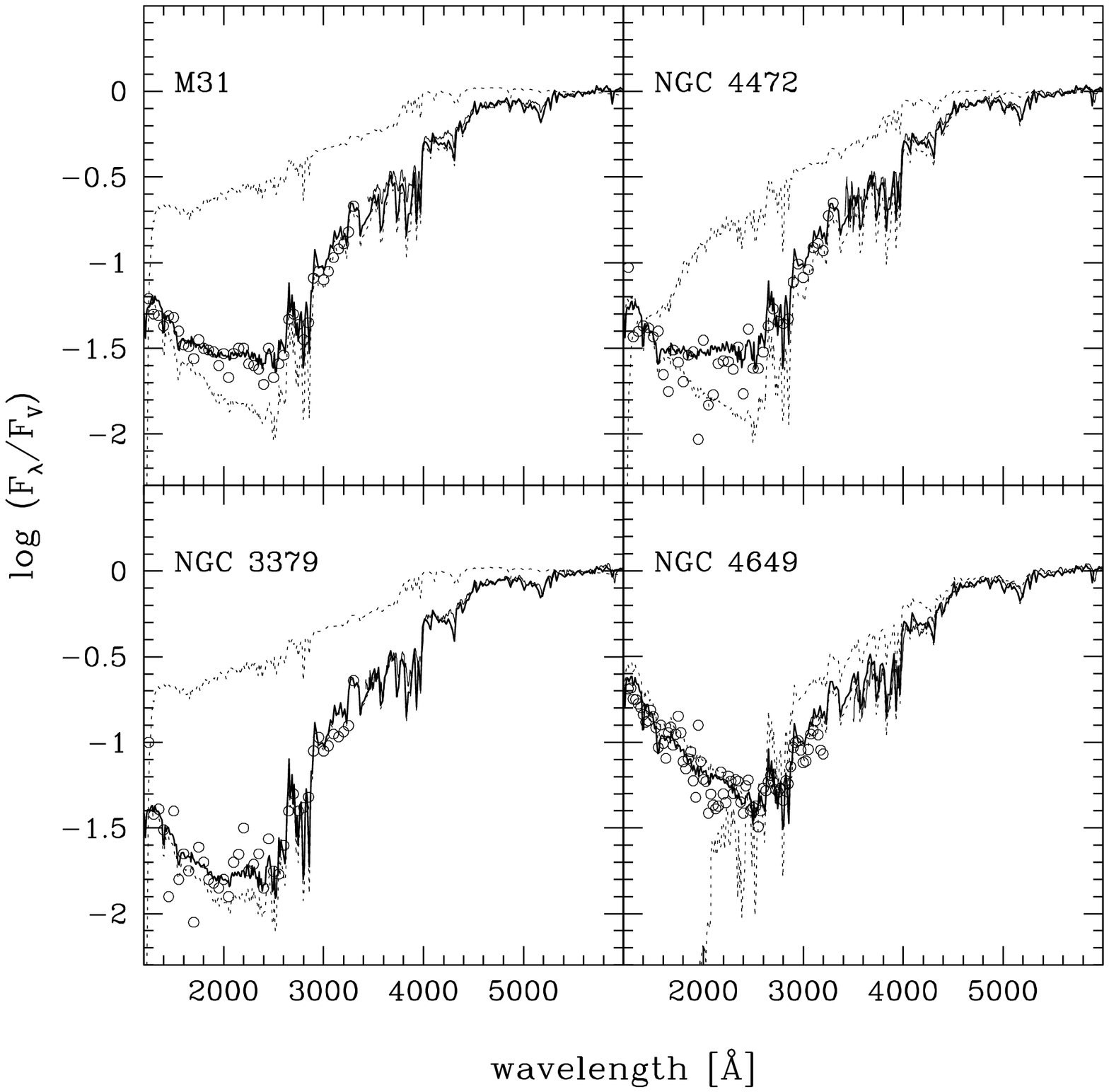,width=\linewidth}
\end{minipage}
\end{center}
\caption{Spectral energy distributions as a function of wavelength.
{\em IUE} data for $\lambda \leq 3300$~\AA\ (circles) are from Burstein et
al.\ 1988 and L.\ Buson (1999, private communication); the spectro-photometry
for $\lambda \geq 3440$~\AA\ (thin solid lines) is from D.\ Hamilton (1999,
private communication). The models from Table~\ref{modeltab} are shown as
thick solid lines, dotted lines are the SEDs of the metal-poor and the
metal-rich components (100 per cent), respectively.}
\label{sed}
\nocite{GR90,Betal88}
\end{figure*}

The fits to the observed SEDs are shown in Fig.~\ref{sed} as thick solid
lines, the dotted lines are the SEDs of the metal-poor and the metal-rich
components. The spectra in the UV (open symbols) are from Burstein et al.\
(1988)\nocite{Betal88} and L.\ Buson (1999, private communication), the
spectro-photometry in the optical (thin solid lines) is from D.~Hamilton
(1999, private communication). In all cases, the observed spectra are well
reproduced in the full wavelength range $\lambda=1200-6000$~\AA. Thus the
amount of metal-poor stars required to reproduce the observed \Hb\ (see
Table~\ref{modeltab}) is consistent with the fluxes in the mid-UV, the model
spectra perfectly match the minimum around 2500~\AA. In the following we
briefly discuss the individual galaxies.

\subsubsection{M31}
\label{m31sec}
As shown in Table~\ref{modeltab} we reproduce the observables of M31 (for
the colors see NGC~4472) with $\Zsub=0.005~\Zsun$ contributing 3 per cent to
the total mass of the population, which corresponds to 6 per cent in $V$, 9
per cent at \Hb\ and 61 per cent at $\lambda=2500$~\AA. Wu et al.\
(1980)\nocite{Wetal80}, instead, claim that not more than 50 per cent
contribution from a metal-poor component at $\lambda=2500$~\AA\ is
compatible with the mid-UV flux of M31. The reason for this ostensible
discrepancy is that the authors fix the metallicity of the metal-rich
component to $\Zmaj=0.5~\Zsun$, which is lower than the value adopted here
by more than a factor of 3 (see Table~\ref{modeltab}). As the SED of a more
metal-rich population has lower flux in the mid-UV, our composite model
requires a larger contribution from the metal-poor component. Using exactly
the same prescriptions as these authors, we are consistent with their
result. Due to the low metallicity of the metal-rich component, however, the
model of Wu et al.\ (1980)\nocite{Wetal80} fails in reproducing the other
observational constraints of M31: it results in colors that are too blue
($\BK=3.48$, $\UV\approx 1.21$), Fe indices that are too low (${\rm
Fe5335}\approx2.01$), and Balmer line strengths that are too high
($\Hb=2.07$).

\subsubsection{NGC 4472}
The Virgo elliptical NGC~4472 has virtually the same properties as M31,
wherefore it can be modeled with the same set of populations. The model
given in the second line of Table~\ref{modeltab} is an alternative option
for both objects assuming a higher contribution from the metal-poor
subcomponent and a higher metallicity of the major population. As a
consequence, the contribution from the low-metallicity population increases
to 70 per cent at 2500~\AA\ and to 12 per cent in $V$. Wu et al.\
(1980)\nocite{Wetal80} derive a lower contribution in $V$ relative to
2500~\AA\ (50 per cent at 2500~\AA\ and 6 per cent in $V$), because the
authors chose the globular cluster M13 (NGC~6205) as a representative of the
metal-poor population. As discussed in Section~\ref{sedcalsec}, this
particular globular cluster has an extreme blue horizontal branch morphology
and exhibits very high fluxes in the mid-UV (see Fig.~\ref{sedcal}), which
implies a lower $V$ over 2500~\AA\ flux ratio.

\subsubsection{NGC 3379}
NGC~3379 exhibits equally high \Hb, but bluer colors and
lower metallic indices. The best fitting model thus has a lower $\Zmaj=\Zsun$
and requires only 1 per cent of $\Zsub=0.005~\Zsun$.

\subsubsection{NGC 4649}
NGC~4649 has the strongest UV upturn (bluest \UVV),
requiring a factor of $\sim 10$ more fuel at late evolutionary phases
than the other cases (see column~(11) in Table~\ref{modeltab}). Given the
low \Hb\ line strength, the best fit is obtained with a model
containing only old metal-rich populations ($\Zsub=0.3~\Zsun$). This
object strongly supports the view that the UV rising branch and \Hb\
line strength are not produced by the same kind of stars.

\subsection{Models for $\Hb>2$ \AA}
The model grid in Fig.~\ref{fehb} extends to $\Hb\approx 2$~\AA, while some
low-mass field ellipticals exhibit stronger Balmer lines. In principle such
high \Hb\ can be obtained by increasing the weight of the low-metallicity
component. For instance, the compact dwarf elliptical M32 (NGC~221) with
$\Hb=2.37\pm 0.12$, ${\rm Fe5335}=2.54\pm 0.20$ (Kormendy \& Bender
1999)\nocite{KB99} can be modeled by the 14-Gyr old two-component population
with $\Zmaj=2-3~\Zsun$ and 20 per cent of $\Zsub=0.005~\Zsun$. The SED of
this model with such a large fraction of very metal-poor stars, however is
not compatible with the fluxes observed in the mid-UV (see also Burstein et
al.\ 1984; Rose \& Deng 1999)\nocite{Betal84,RD99}. On the other hand, from
a deep {\em HST} color magnitude diagram, Grillmair et al.\
(1996)\nocite{Getal96} conclude that there is no evidence for an
intermediate age population ($t\la 2$~Gyr). It is worth mentioning, however,
that M32 is a peculiar object that does not have the properties of the
average elliptical galaxy population.

Alternatively to the addition of metal-poor stars, metal-rich populations
with blue horizontal branch morphologies could reproduce strong \Hb\ without
invoking young ages. Such hot horizontal branch stars are indeed observed in
the metal-rich globular clusters NGC~6388 and NGC~6441 of the galactic bulge
(Rich et al.\ 1997)\nocite{Retal97} and in M32 (Brown et al.\
2000)\nocite{Betal00}. A possible mechanism to produce these temperatures is
enhanced mass loss along the RGB evolutionary phase. The investigation of
this channel to obtain strong \Hb\ in old metal-rich populations will be the
subject of a future paper.

\section{Discussion}
\label{discusssec}
\subsection{Metallicity distributions in ellipticals}
The principle idea of this paper is to enhance the Balmer line strengths of
old metal-rich stellar populations with a small fraction of metal-poor
stars. The real metallicity distribution of the stellar populations in
elliptical galaxies is difficult to assess observationally, as the stars
cannot be resolved. The {\em HST} color magnitude diagram of M32 (Grillmair
et al.\ 1996)\nocite{Getal96} and the spectroscopy of K giants in the Bulge
(Rich 1988)\nocite{Ri88} show that these spheroids contain a tail of low
metallicity stars. The closest giant elliptical for which color magnitude
diagrams are available is NGC~5128 (Harris, Harris \& Poole
1999)\nocite{HHP99}. Analyzing deep {\em HST} images in the outer halo of
NGC~5128, the authors find a differential metallicity distribution that is
well reproduced by two closed-box-like chemical enrichment scenarios
implying a large number of extremely metal-poor stars. In particular the
differential shape of the distribution implies a formation picture in which
a metal-poor and a metal-rich population are formed separately from each
other.

This picture gets further support from a number of studies that discover
bimodal color and metallicity distributions of globular clusters in at least
half of the early-type galaxy population (Zepf \& Ashman 1993; Gebhardt \&
Kissler-Patig 1999)\nocite{ZA93,GK99}. Spectroscopic and photometric
investigations indicate that both populations are old (Cohen, Blakeslee \&
Ryzhov 1998; Kissler-Patig et al.\ 1998; Kissler-Patig, Forbes \& Minniti
1998; Kundu et al.\ 1999; Puzia et al.\
1999)\nocite{Kietal98,CBK98,KFM98,Kuetal99,Puetal99}. More specifically,
Puzia et al.\ (1999)\nocite{Puetal99} find that the two populations in
NGC~4472 are old and coeval, with metallicities $\sim 0.05~\Zsun$ and
$\sim\Zsun$. In the framework of hierarchical structure formation galaxies
are built by mergers of smaller objects (e.g., White \& Rees 1978; White \&
Frenk 1991; Kauffmann, White \& Guiderdoni 1993)\nocite{WR78,WF91,KWG93}. A
merger of coeval systems without newly induced star formation would result
in a coeval composite population. The accretion of a dwarf galaxy by a
larger system (minor merger) qualitatively explains the existence of a
metal-poor subpopulation.

There is an additional important effect that is independent of the assumed
scheme of galaxy formation. As discussed by Greggio (1997)\nocite{Gre97}, a
sizable fraction of metal-poor stars from the outer parts of a galaxy
contaminate the light coming from the center due to projection and orbital
mixing (Ciotti, Stiavelli \& Braccesi 1995)\nocite{CSB95}. The authors find
that projection effects lower the actual central metallicity by roughly 10
per cent. The metal-poor components in the composite models of M31, NGC~4472
and NGC~3379 (Table~\ref{modeltab}) dilute the metallicities of the
metal-rich populations by 3, 6, and 4 per cent, respectively. Projection
effects alone thus may be sufficient to explain the amount of metal-poor
stars considered in the present analysis.

\subsection{Evidence against recent star formation}
There are observational indications from the Fundamental Plane (Djorgovski
\& Davis 1987; Dressler et al.\ 1987; Bender, Burstein \& Faber 1992, 1993;
Renzini \& Ciotti 1993)\nocite{DD87,Dretal87,BBF92,BBF93,RC93} and the
color-magnitude relation (Bower, Lucey \& Ellis 1992)\nocite{BBF92} that
limit the fraction of the young population to less than 10 per cent. As late
star formation leads to low $\alpha$/Fe ratios (Thomas, Greggio \& Bender
1999; Thomas 1999)\nocite{TGB99,Th99a}, the addition of a young population
provokes inconsistencies with the high $\alpha$/Fe ratios observed in
elliptical galaxies (e.g., Worthey et al.\ 1992)\nocite{WFG92}. Also the
redshift evolution of colors (Arag\'on Salamanca et al.\
1993)\nocite{Aetal93}, of the Mg-$\sigma$ relation (Bender, Ziegler \&
Bruzual 1996)\nocite{BZB96}, of the Kormendy relation (Ziegler et al.\
1999)\nocite{Zieetal99}, and of the color gradients in cluster ellipticals
(Saglia et al.\ 2000)\nocite{Saetal00} leave little space for a significant
contribution from a young subcomponent. The addition of an old, metal-poor
population as discussed in this paper can reconcile these findings with
strong Balmer lines.

Disturbed field ellipticals that have strong \Hb\ absorption (Schweizer et
al.\ 1990)\nocite{Schetal90} and blue optical colors (Schweizer \& Seitzer
1992)\nocite{SS92} do not show any signature of recent star formation
activity in the infrared colors (Silva \& Bothun 1998a,
1998b)\nocite{SB98a,SB98b}. As discussed by these authors, recent mergers
may not have been accompanied by significant star formation, but metallicity
effects are favored to explain the enhanced
\Hb\ line strengths.

\section{Summary}
\label{sumsec}
In this paper we show that strong Balmer lines can be produced by composite
populations that contain a small fraction of old {\em metal-poor} stars.

The key to this result is that at low metallicities and old ages our SSP
models show a steep increase of \Hb\ with decreasing metallicity, in
accordance to what is observed in metal-poor globular clusters. We compute
composite stellar population models that consist of an underlying old
metal-rich population and a small fraction of an old metal-poor
subcomponent. Assuming a 10 per cent contribution from metal-poor stars, we
construct a grid of models in the \Hb-Fe5335 and the \Hb-\BK\ planes that
reproduce \Hb\ up to $2$~\AA\ and covers most of the elliptical galaxy data.
It should be emphasized that the same conclusion holds for all Balmer lines,
e.g.\ H${\gamma}$ and H${\delta}$. More specifically, the data of all
cluster and luminous field ellipticals can be explained by our models
without invoking young ages. The scatter in Balmer line strengths is then
caused by a spread in the metallicity and/or the weight of the metal-poor
component. Most ellipticals can be modeled with less than 10 per cent of
metal-poor stars as the data scatter about $\Hb\sim 1.6$~\AA.

We further show that the spectral energy distributions of the
composite models that reproduce the indices and colors of
representative examples agree well with the observed spectra in the
wavelength range $1200-6000$~\AA. Our models are perfectly compatible
with the observed minimum at 2500~\AA, that tightly constrains the
possible amount of metal-poor stars.

We conclude that the age-metallicity degeneracy for complex systems like
elliptical galaxies still remains to be solved. The perturbation with old
{\em metal-poor} stars to obtain strong Balmer lines is alternative to the
addition of a {\em young} metal-rich population (e.g., de Jong \& Davies
1997; Kuntschner 2000; Longhetti et al.\ 2000)\nocite{DD97,Ku00,Letal00}.
The key to discriminate between the two options lies in the evolution of
Balmer lines with redshift. A young population of $2-5$~Gyr should leave its
fingerprints with a significant peak of Balmer line strengths at redshifts
$z\sim 0.15-0.4$, depending on the cosmology, which has not been detected so
far (Ziegler \& Bender 1997)\nocite{ZB97}. The model presented here,
instead, predicts Balmer lines to become monotonically stronger with
redshift.

\acknowledgments
We are grateful to S.\ Cassisi, D.\ Hamilton, D.\ Mehlert, and L.\ Buson,
for providing electronic versions of models and data. We thank R.\ Bender
and M.\ Kissler-Patig for the careful reading of the manuscript. For many
helpful discussions, we thank N.\ Drory, L.\ Greggio, U.\ Hopp, and R.\
Saglia. The referee is acknowledged for the very useful comments that
improved the first version of the paper. This work was supported by the
"Sonderforschungsbereich 375-95 f\"ur Astro-Teilchenphysik" of the Deutsche
Forschungsgemeinschaft.

\end{document}